\documentclass[letterpaper,english,aip,pop,reprint]{revtex4-1}
\usepackage{lmodern}

\usepackage[T1]{fontenc}
\usepackage[latin9]{inputenc}
\usepackage{bm}
\usepackage{amsmath}
\usepackage{amssymb}
\usepackage{graphicx}

\makeatletter

\usepackage{bm}

\newcommand{\lilE}{{\scriptscriptstyle E}}
\newcommand{\lilG}{{\scriptscriptstyle G}}
\newcommand{\ab}{\allowbreak}
\newcommand{\nob}{\nobreak}
\newcommand{\ip}{I_{p}}
\newcommand{\btd}{b_{T}}
\newcommand{\bt}{B_{T}}
\newcommand{\qq}{q}
\newcommand{\pa}{\theta}
\newcommand{\tahat}{\hat{\ta}}
\newcommand{\gl}{\nabla_{\parallel}}
\newcommand{\curv}{\mathcal{K}}
\newcommand{\curvx}{\curv^{x}}
\newcommand{\go}{\Gamma_{0s}}
\newcommand{\gn}{\Gamma_{1s}}
\newcommand{\dls}{D_{\parallel s}}
\newcommand{\zon}[1]{\left\langle #1\right\rangle }
\newcommand{\zonin}[1]{\langle#1\rangle}
\newcommand{\xx}{x}
\newcommand{\zs}{Z}
\newcommand{\ms}{m_{s}}
\newcommand{\tso}{T_{s0}}
\newcommand{\ns}{n_{s}}
\newcommand{\nso}{n_{s0}}
\newcommand{\ninz}{\tilde{n}_{i}}
\newcommand{\ues}{\bm{u}_{\lilE s}}
\newcommand{\ueix}{u_{\lilE i}^{x}}
\newcommand{\ueip}{u_{\lilE i\pa}}
\newcommand{\ueit}{u_{Ei}^{(2)}}
\newcommand{\uls}{u_{\parallel s}}
\newcommand{\rs}{\rho_{s}}
\newcommand{\pnz}{\tilde{\phi}}
\newcommand{\pg}{\phi_{\lilG}}
\newcommand{\gni}{\Gamma_{i}}
\newcommand{\guli}{\Pi_{\parallel}}
\newcommand{\gue}{\Pi_{E}}
\newcommand{\nul}{\nu_{\parallel}}
\newcommand{\tangm}{L_{\ta}}
\newcommand{\om}{\omega}
\newcommand{\Ens}{E_{ns}}
\newcommand{\Ensin}{E_{i}^{\mathrm{s}}}
\newcommand{\Els}{E_{\parallel s}}
\newcommand{\Elic}{E_{\parallel}^{\mathrm{c}}}
\newcommand{\Eue}{E_{\lilE}}
\newcommand{\Euez}{\Eue^{\mathrm{z}}}
\newcommand{\nsin}{\nni^{\mathrm{s}}}
\newcommand{\ulcos}{u_{\parallel}^{\mathrm{c}}}
\newcommand{\uezon}{u_{{\scriptscriptstyle \negthinspace E}}^{\mathrm{z}}}
\newcommand{\tbo}{T_{a}}
\newcommand{\tmflux}{\Pi_{\ta}}
\newcommand{\tmfluxg}{\tmflux^{\mathrm{(G)}}}
\newcommand{\tmfluxt}{\Pi_{\ta}^{\mathrm{(2)}}}
\newcommand{\eavg}[1]{\overline{#1}}

\newcommand{\vts}{v_{ts}}
\newcommand{\ocs}{\Omega_{cs}}
\newcommand{\bpd}{b_{p}}
\newcommand{\spa}{\vartheta}
\newcommand{\flv}{\rho}
\newcommand{\sta}{\xi}
\newcommand{\zz}{z}
\newcommand{\flvh}{\hat{\flv}}
\newcommand{\phat}{\hat{p}}
\newcommand{\sumsp}{\sum_{s}}
\newcommand{\dx}{d\mathcal{V}}
\newcommand{\rip}{\rho_{i\pa}}
\newcommand{\ocip}{\Omega_{ci\pa}}
\newcommand{\Qe}{Q_{e}}
\newcommand{\Qi}{Q_{i}}
\newcommand{\Lp}{L_{p}}
\newcommand{\fl}{f_{L}}
\newcommand{\lpe}{L_{pe}}
\newcommand{\lpi}{L_{pi}}
\newcommand{\bp}{B_{p}}
\newcommand{\bom}{B_{0}}
\newcommand{\mr}{r}
\newcommand{\Mr}{R}
\newcommand{\Ro}{\Mr_{0}}
\newcommand{\ta}{\zeta}
\newcommand{\pta}{\partial_{\ta}}

\newcommand{\pahat}{\hat{\pa}}
\newcommand{\mi}{m_{i}}
\newcommand{\teo}{T_{e0}}
\newcommand{\tio}{T_{i0}}
\newcommand{\dli}{D_{\parallel i}}
\newcommand{\px}{\partial_{\xx}}
\newcommand{\nni}{n_{i}}
\newcommand{\nne}{n_{e}}
\newcommand{\nio}{n_{i0}}
\newcommand{\neo}{n_{e0}}
\newcommand{\uei}{\bm{u}_{\lilE i}}
\newcommand{\uli}{u_{\parallel i}}

\newcommand{\bhat}{\hat{b}}
\newcommand{\ri}{\rho_{i}}
\newcommand{\ptot}{\phi}
\newcommand{\vti}{v_{ti}}
\newcommand{\oci}{\Omega_{ci}}
\newcommand{\pspa}{\partial_{\spa}}
\newcommand{\psta}{\partial_{\sta}}

\newcommand{\Eli}{E_{\parallel i}}

\newcommand{\nenz}{\tilde{n}_{e}}
\newcommand{\tim}{t}
\newcommand{\pt}{\partial_{\tim}}
\newcommand{\gp}{\nabla_{\perp}}
\newcommand{\plap}{\nabla_{\perp}^{2}}
\newcommand{\ev}{\bm{E}}
\newcommand{\bv}{\bm{B}}
\newcommand{\exb}{\ev\times\bv}
\newcommand{\kp}{k_{\perp}}
\newcommand{\kl}{k_{\parallel}}
\newcommand{\Bm}{B}
\newcommand{\ee}{e}

\makeatother

\usepackage{babel}
\begin{document}

\title{Parasitic Momentum Flux in the Tokamak Core}

\author{T. \surname{Stoltzfus-Dueck}}

\affiliation{Princeton University, Princeton, NJ 08544}

\email{tstoltzf@princeton.edu}

\date{\today}
\begin{abstract}
A geometrical correction to the $\bm{E}\times\bm{B}$ drift causes
an outward flux of cocurrent momentum whenever electrostatic potential
energy is transferred to ion parallel flows. The robust symmetry breaking
follows from the free energy flow in phase space and does not depend
on any assumed linear eigenmode structure, acting both for axisymmetric
fluctuations (such as geodesic acoustic modes) as well as more general
nonaxisymmetric fluctuations. The resulting rotation peaking is countercurrent
and scales as electron temperature over plasma current. This peaking
mechanism can only act when fluctuations are low-frequency enough
to excite ion parallel flows, which may explain some recent experimental
observations related to rotation reversals.
\end{abstract}

\pacs{52.25.Dg, 52.25.Fi, 52.25.Xz, 52.30.Gz, 52.35.We, 52.55.Dy,
52.55.Fa}


\maketitle

Tokamak plasmas without applied torque routinely rotate spontaneously
in the toroidal (symmetry) direction, exhibiting nonzero, sheared
toroidal rotation profiles. \citep{deGrassie09rev} This so-called
``intrinsic'' rotation is not only of fundamental interest: toroidal
rotation helps suppress certain instabilities\citep{Strait95}  and
its shear may reduce turbulent heat transport.\citep{Biglari90} These
advantages are important for future burning plasma devices such as
ITER, in which the dominant $\alpha$-heating will not exert toroidal
torque, unlike the neutral beam heating typical of present-day devices.
\citep{Doyle07}

Although experimentally measured intrinsic rotation profiles are very
diverse, many exhibit three distinct radial regions: an edge region
with cocurrent rotation (toroidal rotation in the direction of the
plasma current $\ip$), a mid-radius ``gradient region'' where rotation
either becomes increasingly countercurrent with decreasing radius
(countercurrent peaking) or stays relatively flat, and a flat or weakly
cocurrent-peaked central region affected by sawtoothing.\cite{deGrassie07,*Eriksson09,*McDermott11mom,StoltzfusDueck15,*StoltzfusDueck15rotpop,Rice11nf,Angioni11,Sauter10iaea}
Previous theoretical,\cite{StoltzfusDueck12,*StoltzfusDueck12pop}
numerical,\citep{Seo14} and experimental\citep{StoltzfusDueck15,StoltzfusDueck15rotpop}
work suggests that the edge rotation is driven by the interaction
of passing-ion drift orbit excursions with spatial variation of the
turbulent fluctuations. The present work focuses on the ``gradient
region'' at intermediate radius, where radial variation of plasma
parameters is much slower, allowing other effects to compete with
those of orbit excursions. 

Over the last decade, intrinsic rotation at mid-radius has undergone
intense theoretical and experimental investigation. Nonaxisymmetric
magnetic fields can strongly affect the toroidal rotation.\citep{Park13}
The present work will focus exclusively on the case of axisymmetric
confining magnetic field, for which the conservation of toroidal
angular momentum\cite{Scott10mom,*Brizard11} excludes the possibility
of a self-generated torque. Intrinsic rotation must therefore result
from a nondiffusive component to the momentum flux. Neoclassical (collisional)
momentum fluxes are much too small to explain experimental observations,
implying that turbulent transport is dominant.\citep{deGrassie09rev}
A number of turbulent calculations suggest the presence of a momentum
pinch, a component of momentum flux that is proportional to the toroidal
rotation itself, rather than its gradient.\cite{Hahm07,*Peeters09cor}
However, these models cannot explain the common observation of  sheared
velocity profiles passing through zero.\citep{deGrassie07,Eriksson09,Sauter10iaea,Rice11nf,Angioni11,McDermott11mom,StoltzfusDueck15rotpop}
Such measurements imply the presence of a ``residual stress,'' meaning
a momentum flux contribution that is independent of both toroidal
rotation and its radial gradient. For up-down symmetric geometries,
often a good approximation for tokamak core plasmas, symmetry arguments
restrict the leading-order momentum flux terms from driving residual
stress.\citep{Peeters11} Theoretical work has accordingly focused
on  symmetry-breaking mechanisms\citep{Peeters11} such as $\exb$
shear,\citep{Dominguez93} up-down-asymmetric geometry,\citep{Camenen09}
and polarization effects.\citep{McDevitt09pop} Particularly challenging
to theory are the experimental observations of rotation reversals
in the ``gradient region,'' in which countercurrent rotation peaking
 suddenly flattens or switches to weak cocurrent peaking when plasma
density or current cross threshold values.\citep{Sauter10iaea,Angioni11,Rice11nf}
The rapidity of these reversals suggests that the direction of peaking
is not determined by neoclassical flows or other quantities that vary
smoothly with plasma parameters, following instead from the properties
of the turbulence itself, which may suddenly change character e.g.~as
an instability threshhold is crossed. In this letter, I identify
a geometrical correction to the $\exb$ drift, neglected in all previous
analytical work, that causes the free energy flows within the turbulence
to drive a robust, fully nonlinear symmetry-breaking momentum flux.
This flux causes counter-current core rotation peaking consistent
with experimental measurements, and explains several observations
related to rotation reversals.  

\begin{figure}
\noindent \begin{centering}
\includegraphics[clip,width=3.35in]{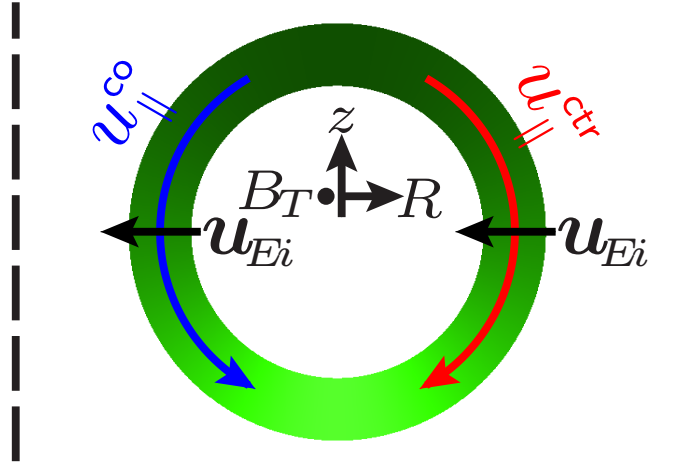}
\par\end{centering}

\protect\caption{\label{fig:gam_flux_cartoon}Poloidal cut of a low-frequency axisymmetric
fluctuation, with axis of symmetry on the left. Darker shading shows
larger $\ninz$, proportional to $\pnz$ by the low-$\kp$ electron
adiabatic response. Given enough time, ions flow out of the dense
region along the magnetic field, causing counter-current (red) toroidal
flow toward decreasing $\pa$ and co-current (blue) toward increasing
$\pa$. The poloidal variation of $\pnz$ causes an $\exb$ flow that
is inward for the countercurrent ion flux and outward for the cocurrent
flux. Reversing the toroidal magnetic field switches the poloidal
direction of counter- and cocurrent flow as well as the sign of the
$\exb$ drift $\uei$, leaving the momentum flux unchanged. The poloidal
orientation of the density perturbation has no effect on the sign
or magnitude of this momentum flux.}
\end{figure}

To develop intuition, consider first a low-frequency axisymmetric
density perturbation, as sketched in Fig.~\ref{fig:gam_flux_cartoon}.
At low frequencies and large scales, electron parallel force balance
ensures that the nonzonal electrostatic potential $\pnz$ is proportional
to the nonzonal  ion gyrocenter density $\ninz$. The pressure gradient
and electric field then cause ions to flow out of the dense region
along the magnetic field. The poloidal electric field also causes
a radial $\exb$ drift that advects counter- (co-)current ion momentum
inward (outward), regardless of the signs of $\ip$ and the toroidal
magnetic field $\bt$. 

Key to this mechanism is a dual role for the weak electric field caused
by the poloidal variation of the potential $\ptot$ on length scales
comparable to the minor radius $\mr$. The nonvanishing parallel component
of this electric field allows it to cause local ion acceleration,
resulting in energy transfer between electrostatic potential ($\ptot$)
and parallel ion flow ($\uli$). Because the background plasma gradients
predominantly supply energy to even moments of the distribution function
(such as density), while odd moments (such as $\uli$) are subjected
to dissipation,\citep{Scott10en,StoltzfusDueck16rotpop} steady-state
energy balance often requires a net transfer of free energy from the
potential (a function of even moments) to the ion parallel flows,
causing a statistical symmetry breaking in the corresponding energy
transfer term. Although toroidal angular momentum conservation
does not allow the self-generated electric field to impart a net torque
to the plasma, the weak radial $\exb$ drift due to the poloidally
varying $\ptot$ may transport toroidal angular momentum in the radial
direction. The correlations between the ion parallel flows and the
weak radial $\exb$ drift, resulting from the statistical symmetry
breaking due to energy transfer, cause this part of the momentum flux
to have a preferred sign, independent of plasma rotation and its radial
gradient. In this letter, we will consider this residual stress in
two separate cases: first a simpler special case with axisymmetric
fluctuations, where the momentum flux occurs due to damping of geodesic
acoustic modes (GAMs) via ion parallel flows, and later a more general
case including nonaxisymmetric fluctuations, where the momentum flux
can occur for any turbulent fluctuations in which energy transfer
from potential to ion parallel flow is nonnegligible.

Both calculations use the simplest model capturing the relevant physics:
the large-aspect-ratio limit of the electrostatic, isothermal gyrofluid
equations in a  radially thin geometry,\citep{Scott03,Scott10en}
written in cgs units as
\begin{align}
\pt\ns\negthickspace+\negthickspace\ues\negthickspace\cdot\negthickspace\nabla\negthinspace\negthinspace\left(\ns\negthickspace+\negthickspace\nso\right)\negthinspace= & \curv\Bigl(\negthinspace\frac{\ns\tso}{\zs\ee}\negthickspace+\negthickspace\nso\pg\negthinspace\Bigr)\negthickspace-\negmedspace\nso\gl\uls,\negthickspace\negthickspace\label{eq:ns_eqn}\\
\ms\nso\negthickspace\left(\pt\negthickspace+\negthickspace\ues\negthickspace\cdot\negthickspace\nabla\right)\negthinspace\uls\negthickspace= & -\gl\left(\ns\tso+\zs\ee\nso\pg\right)\nonumber \\
 & +\negthickspace\ms\nso\negthinspace\Bigl[\frac{2}{\zs\ee}\tso\curv(\uls)\negthickspace-\negthickspace\dls\Bigr]\negthinspace,\negthickspace\negthickspace\label{eq:uls_eqn}\\
\sumsp\negthickspace\nso\zs^{2}\ee^{2}\frac{1\negthickspace-\negthickspace\go}{\tso}\ptot= & \sumsp\zs\ee\gn\ns,\label{eq:gyroflu_poiss}
\end{align}
with species subscript $s$ meaning ions $i$ or electrons $e$;
species charge state $\zs$ (-1 for electrons), mass $\ms$, and (constant)
temperature $\tso$; fluctuating $\ns$ and equilibrium $\nso$ species
density (assuming $\zs\nio=\neo$); $\exb$ drift $\ues\doteq(c/\Bm)\bhat\times\nabla\pg$
; gyroaveraged potential $\pg\doteq\gn\ptot$; curvature operator
$\curv\doteq-(2c/\Bm)\bhat\times\nabla\ln\Bm\cdot\nabla$  capturing
the magnetic drifts and $\exb$ divergence; parallel gradient $\gl\doteq\bhat\cdot\nabla$;
parallel flow velocity $\uls$; and dissipation operator $\dls$.
The gyroaveraging operators $\go$ and $\gn$ take the low-$\kp$
limits $(1-\go)\to-\rs^{2}\plap$ and $\gn\to1$, for $\rs=\vts/|\ocs|$
the species gyroradius, with thermal speed $\vts\doteq(\tso/\ms)^{1/2}$
and gyrofrequency $\ocs\doteq\zs\ee\bom/\ms c$. We take safety factor
$\qq=\bt\mr/\bp\Ro$ order unity, so the poloidal field and inverse
aspect ratio are comparably small, $\bp/\bt\sim\mr/\Ro\ll1$, allowing
explicit appearances of $\Bm$ and $\Mr$ to be replaced with representative
constants $\bom$ and $\Ro$, and setting $\btd\doteq\bhat\cdot\tahat\to\pm1$
for magnetic and toroidal directions $\bhat$ and $\tahat$. Since
the toroidal component of $\ues$ is small in $\bp/\bt$, Eqs.~\eqref{eq:ns_eqn}--\eqref{eq:gyroflu_poiss}
conserve a simplified toroidal angular momentum involving only the
zonal (flux-surface) average $\zonin{\cdots}$ of $\uli$, assuming
$\zonin{\dli}=0$:
\begin{equation}
\pt\zon{\tangm}=-\px\zon{\tmflux},\label{eq:gyroflu_momcons}
\end{equation}
with toroidal angular momentum density and flux $\tangm\doteq\mi\nio\btd\Ro\uli$,
$\tmflux\doteq\mi\nio\btd\Ro(\ueix-\frac{2\tio}{\zs\ee}\curvx)\uli$,
with $\ueix\doteq\uei\cdot\nabla\xx$, $\curvx\doteq\curv(\xx)$,
and radial (flux-surface) label $\xx$. Eq.~\eqref{eq:gyroflu_momcons}
shows that toroidal angular momentum is advected by the $\exb$ and
magnetic drifts, without sources or sinks.

A nondiffusive momentum flux as in Fig.~\ref{fig:gam_flux_cartoon}
may be driven by  geodesic acoustic mode (GAM) damping, which we
may treat in a shearless simple-circular geometry, $\curvx\to(2c\btd/\bom\Ro)\sin\pa$
for poloidal angle $\pa$. Following Ref.~\onlinecite{Scott05njp},
we retain only one axisymmetric Fourier component from each of Eqs.~\eqref{eq:ns_eqn}--\eqref{eq:gyroflu_poiss},
specifically $\nsin\doteq\zonin{\nni\sin\pa}$, $\ulcos\doteq\zonin{\uli\cos\pa}$,
and $\uezon\doteq\zonin{\ueip}$ with $\ueip\doteq\uei\cdot\pahat=\btd\frac{c}{\bom}\px\ptot$.
We neglect electron polarization and take low-$\kp$ gyroaveraging
and quasineutrality $\zs\zonin{\nni\sin\pa}\approx\zonin{\nne\sin\pa}$,
and electron adiabatic response $\teo\zonin{\nne\sin\pa}\approx\ee\neo\zonin{\ptot\sin\pa}$,
obtaining
\begin{align}
\pt\nsin & =\nio\uezon/\Ro+\bpd\nio\ulcos/\mr-\px\zon{\gni\sin\pa},\label{eq:nsin}\\
\mi\nio\pt\ulcos & =\negthinspace-\bpd\tbo\nsin/\mr\negthinspace-\negthinspace\nul\mi\nio\ulcos-\negthinspace\px\negthinspace\zon{\guli\cos\pa}\negthinspace,\negthickspace\label{eq:ulicos}\\
\nio\mi\pt\uezon & =-2\tbo\nsin/\Ro-\px\zon{\gue},\label{eq:uezon}
\end{align}
in which $\bpd\doteq\bhat\cdot\pahat\ll1$ and $\tbo\doteq\tio+\zs\teo$.
We have taken $\zonin{\dli\cos\pa}\to\nul\ulcos$ for parallel flow
damping rate $\nul$. The $\gni$, $\guli$, and $\gue$ terms respectively
capture the divergences of ion density, parallel/toroidal momentum,
and $\exb$/poloidal momentum fluxes due to unresolved Fourier components.
The $\px\zonin{\gue}$ form follows from Eq.~\eqref{eq:gyroflu_poiss},
with $\gue\approx\mi\nio\ueip\ueix$ plus FLR corrections.\citep{Scott10mom}
Linearizing Eqs.~\eqref{eq:nsin}--\eqref{eq:uezon} and neglecting
$\gni$, $\guli$, and $\gue$ yields a simple dispersion relation
\begin{equation}
\om^{2}-2\frac{\tbo}{\mi\Ro^{2}}=\frac{\om}{\om+i\nul}\frac{\tbo}{\mi\qq^{2}\Ro^{2}}.\label{eq:simp_GAM_disp}
\end{equation}
For $\qq\gg1$, Eq.~\eqref{eq:simp_GAM_disp} contains a pair of
weakly damped high-frequency GAMs $\om\approx\pm(2\tbo/\mi\Ro^{2})^{1/2}-i\nul/4\qq^{2}$.
However, for $\qq$ near 1, as is typical in tokamak core plasmas,
the GAMs damp at a significant fraction  of $\nul\sim\vti/\qq\Mr$,
 as seen in more detailed kinetic calculations.\cite{Novakovskii97,*Sugama06,*Sugama06erratum}

To evaluate and physically understand the resulting toroidal momentum
flux, we examine the free energy balance for $\nsin$, $\ulcos$,
and $\uezon$:
\begin{align}
\negthickspace\pt\Ensin & =2\tbo\nsin[\uezon/\Ro+\bpd\ulcos/\mr-\nio^{-1}\px\zon{\gni\sin\pa}],\label{eq:nsin_en}\\
\negthickspace\pt\Elic & =\negthinspace-2\ulcos(\bpd\tbo\nsin/\mr\negthinspace+\negthinspace\nul\mi\nio\ulcos\negthinspace+\negthinspace\px\negthinspace\zon{\guli\cos\pa}),\negthickspace\label{eq:ulicos_en}\\
\negthickspace\pt\Euez\negthinspace & =\negthinspace-2\tbo\nsin\uezon/\Ro-\uezon\px\zon{\gue},\label{eq:uezon_en}
\end{align}
in which $\Ensin\doteq\tbo(\nsin)^{2}/\nio$, $\Elic\doteq\mi\nio(\ulcos)^{2}$,
and $\Euez\doteq\frac{1}{2}\nio\mi(\uezon)^{2}$. Turbulence simulations
show that the Reynolds stress ($\gue$) typically acts as a source
for $\Euez$, the geodesic transfer term ($\propto\nsin\uezon$) moves
free energy from $\Euez$ to $\Ensin$, and both parallel flow excitation
($\propto\nsin\ulcos$) and the turbulent density flux sideband ($\propto\px\zonin{\gni\sin\pa}$)
move energy out of $\Ensin$.\citep{Miyato04,Scott05njp} Note next
that the electron adiabatic response combined with nonzero $\nsin$
implies a radial $\exb$ drift $\zonin{\ueix\cos\pa}=-c\btd\zonin{\ptot\sin\pa}/\Bm\mr=-c\btd\teo\nsin/\ee\nio\Bm\mr$.
Recalling Eq.~\eqref{eq:gyroflu_momcons}, this beats with $\ulcos$
to cause a contribution $\tmfluxg=-2(\zs\teo/\tio)(\ri/\mr)\mi\Ro\vti\nsin\ulcos$
to the toroidal angular momentum flux $\zonin{\tmflux}$. Since $\tmfluxg$
is directly proportional to the parallel flow excitation term in Eqs.~\eqref{eq:nsin_en}
and~\eqref{eq:ulicos_en}, we conclude that energy transfer from
the pressure sideband to the parallel ion flow necessarily implies
an outflux of cocurrent toroidal angular momentum. Indeed, since the
turbulent flux term $-2\ulcos\px\zonin{\guli\cos\pa}$ will typically
transfer energy out of $\Elic$, we may use the statistical average
$\eavg{\cdots}$ of Eq.~\eqref{eq:ulicos_en}, $-\bpd\tbo\eavg{\nsin\ulcos}/\mr\gtrsim\nul\mi\nio\eavg{(\ulcos)^{2}}$,
to conservatively estimate the (signed) momentum flux as $\tmfluxg\sim2(\nul/\bpd\oci)(\zs\teo/\tbo)\nio\mi\Ro\eavg{(\ulcos)^{2}}$.
Although simple ordering estimates suggest this flux may only drive
 toroidal ion thermal Mach numbers of order $(\ri/\mr)$ times the
ratio of GAM kinetic energy over turbulent fluctuations' kinetic energy,
its fixed relation to the free-energy transfer guarantees robust symmetry
breaking whenever there is strong GAM damping acting via ion parallel
flows. Quantitative evaluation of its magnitude will require numerical
simulation.

What is happening here physically? First, Reynolds stress excites
a poloidal $\exb$ flow. Poloidal variation of $\Bm$ causes a divergence
in the $\exb$ velocity, resulting in up-down-asymmetric density fluctuations,
like those sketched in Fig.~\ref{fig:gam_flux_cartoon}. The resulting
poloidal electric field (due to adiabatic electron response) and ion
pressure gradient jointly excite an ion flow along $\bv$. The net
energy flow from Reynolds stress drive to damping via the ion parallel
flow implies a positive correlation of the poloidal electric field
and poloidal ion flow. The poloidal electric field also causes a weak
radial $\exb$ drift. Due to the pitch of the magnetic field, the
poloidal ion flow along the field corresponds to co- (counter-)current
toroidal flow where the $\exb$ drift points radially outward (inward),
which causes countercurrent rotation peaking.

Energy transfer from nonaxisymmetric potential fluctuations to ion
parallel flows can drive an even stronger toroidal momentum flux,
but in order to understand its origin we must first discuss the field-aligned
magnetic coordinates used in most gyrokinetic formulations: Consider
now an axisymmetric geometry with good nested flux surfaces, but otherwise
arbitrary. Radial position is specified by a flux-surface label $\flv$,
which is axisymmetric and satisfies $\bhat\cdot\nabla\flv=0$. Poloidal
position is specified by a distended but axisymmetric poloidal angle
label $\spa$. The third coordinate $\sta$ is chosen so that $\bhat\cdot\nabla\sta=0$,
letting it label perpendicular position within the flux surface. These
choices are not arbitrary: The definition of $\sta$ implies that
$\bhat\cdot\nabla=(\bhat\cdot\nabla\spa)\pspa$ so $\pspa|_{\flv,\sta}=(\bhat\cdot\nabla\spa)^{-1}\bhat\cdot\nabla$
contains only slow variation. The use of an axisymmetric $\flv$ and
$\spa$ implies that the partial $\psta|_{\flv,\spa}$ is proportional
to a simple toroidal derivative $\tahat\cdot\nabla$, since holding
$\flv$ and $\spa$ fixed is equivalent to holding $\Mr$ and vertical
position $\zz$ fixed. This property has two important implications.
First, appropriate choice of $\sta$ allows toroidal periodicity to
imply simple periodicity in $\sta$. Second, $\psta\propto\tahat\cdot\nabla$
vanishes for any axisymmetric quantity, in particular for equilibrium
plasma parameters and the magnetic geometry. These properties allow
one to construct symmetry arguments that the dominant toroidal angular
momentum flux, due to the $\psta\ptot$ portion of $\uei$, must vanish
in the statistical average for leading-order local gyrokinetic formulations
with up-down symmetric magnetic geometry.\citep{Peeters11} In
contrast, the $\pspa\ptot$ portion of $\uei$, neglected in all previous
analytical works, is unrestricted by the symmetry arguments\citep{Peeters11}
and indeed must break symmetry in the (common) case of net energy
transfer from $\ptot$ to ion parallel flows, as we will derive now.

We begin with the contribution of the higher-order part of the $\exb$
drift, in a simple, geometric way. Defining the radial and poloidal
directions $\flvh\doteq(\nabla\flv)/|\nabla\flv|$ and $\phat\doteq\tahat\times\flvh$,
decompose $\bhat=\btd\tahat+\bpd\phat$. Since $\flvh\times\bhat=(\tahat-\btd\bhat)/\bpd$,
the radial component of the $\exb$ drift is 
\begin{equation}
\uei\cdot\flvh=\negthickspace\frac{c}{\Bm}\bhat\times\negthickspace\nabla\pg\cdot\flvh=\negthickspace\frac{c}{\bpd\Bm}\negthickspace\left(\tahat\cdot\negthickspace\nabla\pg-\btd\bhat\cdot\negthickspace\nabla\pg\right)\cdot\label{eq:ExB_decomp}
\end{equation}
The first term is the leading-order contribution, restricted by symmetry.
The second term does not represent true parallel physics, it simply
cancels the parallel gradient contribution that was included in the
first term, leaving the true $\gp\pg$. Although nominally smaller
than the first term by $\kl/\kp\bpd$, it has symmetry-breaking properties,
as we will  identify in its contribution to $\tmflux$: 
\begin{equation}
\tmfluxt=-(c\mi\nio\Ro/\bpd\bom)\uli\gl\pg.
\end{equation}
For emphasis, $\tmfluxt$ does not represent any effect of parallel
acceleration, it is simply the advection of the parallel portion of
toroidal angular momentum by a small but robustly symmetry-breaking
portion of the $\exb$ drift.

\begin{figure}
\noindent \begin{centering}
\includegraphics[bb=2bp 2bp 344bp 184bp,clip,width=3.35in]{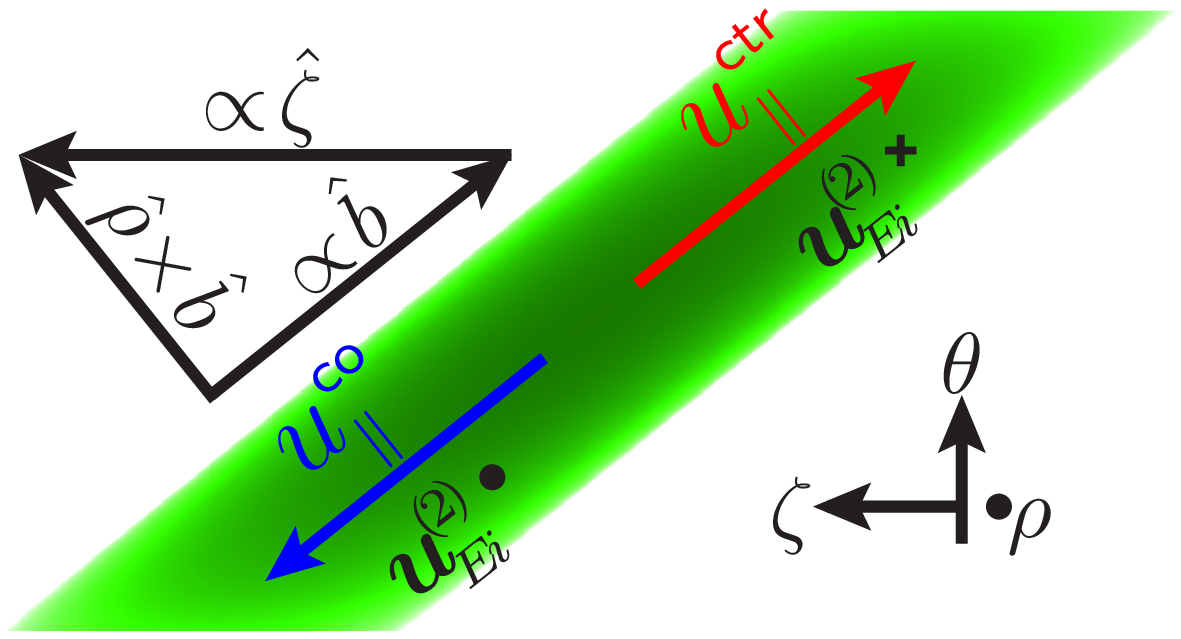}
\par\end{centering}

\protect\caption{\label{fig:dens_cartoon}Side view of a toroidally asymmetric low-frequency
fluctuation, with darker shading again showing larger $\ninz$, proportional
to $\pnz$ by the electron adiabatic response. At frequencies $\lesssim\kl\vti$,
ions flow along $\bhat$ out of the density hump. The symmetry-breaking
portion of the radial $\exb$ drift, $\ueit\protect\doteq-(c\btd/\bpd\Bm)\bhat\cdot\nabla\pg$,
again brings counter- (co-)current parallel momentum inwards (outwards).
Although the $\exb$ drift due to $\pta\pnz$ (not shown) is now nonzero,
its contribution to the momentum flux nearly vanishes by symmetry.}
\end{figure}

To understand the momentum flux caused by $\tmfluxt$ we must examine
the free-energy balance for Eqs.~\eqref{eq:ns_eqn}--\eqref{eq:gyroflu_poiss},
derived following Ref.~\onlinecite{Scott10en}:
\begin{align}
\negthinspace\negthinspace\negthinspace\negthinspace\pt\Ens\negthinspace\negthinspace & =\negthinspace\tso\negthinspace\negthinspace\int\negthinspace\negthinspace\dx\negthinspace\left[\uls\negthinspace\gl\ns\negthinspace+\negthinspace\ns\curv\negthinspace\left(\pg\negthinspace\right)\negthinspace-\negthinspace\ns\negthinspace\frac{\ues\negthinspace\cdot\negthinspace\nabla\nso}{\nso}\right]\negthinspace\negthinspace,\negthinspace\label{eq:ns_en}\\
\negthinspace\negthinspace\negthinspace\negthinspace\pt\Els\negthinspace\negthinspace & =\negthinspace-\negthinspace\negthinspace\negthinspace\int\negthinspace\negthinspace\dx\thinspace\uls\negthinspace\left[\tso\gl\ns\negthinspace\negthinspace+\negthinspace\negthinspace\zs\ee\nso\negthinspace\gl\pg\negthinspace\negthinspace+\negthinspace\ms\nso\dls\right]\negthinspace\negthinspace,\negthinspace\negthinspace\label{eq:uls_en}\\
\negthinspace\negthinspace\negthinspace\negthinspace\pt\Eue\negthinspace\negthinspace & =\negthinspace\sumsp\negthinspace\int\negthinspace\negthinspace\dx\left[-\tso\ns\curv\left(\pg\right)+\zs\ee\nso\uls\gl\pg\right]\negthinspace\negthinspace,\label{eq:poiss_en}
\end{align}
with fluctuating pressure $\Ens\doteq\int\dx\thinspace\frac{1}{2}(\tso/\nso)\ns^{2}$
and parallel flow $\Els\doteq\int\dx\thinspace\frac{1}{2}\ms\nso\uls^{2}$
free energies, $\exb$ energy including FLR corrections $\Eue\doteq\int\dx\thinspace\sumsp\frac{1}{2}(\nso/\tso)\zs^{2}\ee^{2}\ptot\left(1-\go\right)\ptot$,
and volume integral $\int\dx$.   Boundary terms have been assumed
to vanish.  The key point here is that the momentum flux term $\tmfluxt$
is directly proportional to the electrostatic acceleration of ion
parallel flows, $-\zs\ee\nio\int\dx\thinspace\uli\gl\pg$ in  Eq.~\eqref{eq:uls_en}.
Although this term is sometimes referred to as ion Landau damping,
it is in fact conservative, representing a transfer of energy from
the potential $\Eue$ to ion parallel flow $\Eli$. In cases with
damping of turbulence via parallel ion flows, this term will tend
to be positive, so that $\tmfluxt$ transports cocurrent momentum
outward, corresponding to countercurrent rotation peaking,\footnote{For an atypical case with inverse ion Landau damping, meaning energy
transfer from $\Eli$ to $\Eue$, $\tmfluxt$ would reverse sign and
cause cocurrent peaking.} c.\,f.~Fig.~\ref{fig:dens_cartoon}. This will especially occur
when there are density fluctuations at low $\kp$, due to low frequencies,
electron adiabatic response and low-$\kp$ quasineutrality $\ee\neo\pnz\approx\teo\nenz\approx\zs\teo\ninz$,
which reduces $-\int\dx\thinspace\uli[\tio\gl\nni+\zs\ee\nio\gl\pg]\to-\zs\ee\nio\ab(1\nob+\nob\tio/\zs\teo)\int\dx\thinspace\uli\gl\ptot\approx\mi\nio\int\dx\thinspace\uli\dli\ge0$.
Since ion parallel flows are  excited predominantly at low $\kp$,
 we may use this with $\dli\to\nul\uli$ to estimate $\int\dx\thinspace\tmfluxt\approx(\zs\teo/\tbo)(\nul/\ocip)\mi\nio\Ro\int\dx\thinspace\uli^{2}$,
in which $\nul/\ocip\sim\rip/\qq\Ro$ for poloidal ion gyroradius
$\rip\doteq\vti/|\ocip|$, (signed) $\ocip\doteq\zs\ee\bp/\mi c$,
and $\nul\sim\vti/\qq\Ro$.  Alternatively, if a fraction $\fl$
of turbulent free energy is dissipated via ion parallel flows, one
may estimate the resulting volume-averaged momentum flux as $\int\dx\thinspace\tmfluxt\approx\fl\negthinspace(\zs\teo/\tbo)\Ro\mi\vti(\vti/\ocip)[\negthinspace\int\negthinspace\negthinspace\dx(\Qe/\negthinspace L_{pe}\negthinspace+\Qi/\negthinspace L_{pi}\negthinspace)]/\tio$,
with  pressure gradient scale lengths $\lpe=\lpi\doteq\Lp$ and turbulent
radial electron $\Qe$ and ion $\Qi$ heat fluxes. Assuming comparable
turbulent transport coefficients for heat and toroidal angular momentum,
this corresponds to countercurrent velocity peaking with ion thermal
Mach number of order $\fl(\zs\teo/\tio)(\rip/\Lp)$. Although the
isothermal model cannot distinguish between a particle and a heat
flux, the energy balance  for a six-moment non-isothermal gyrofluid
model clearly shows that the necessary density fluctuations may be
driven by an electron or ion heat flux, even in the absence of a particle
flux.\citep{Scott10en} Analogous manipulations in a gyrokinetic
formulation also lead to the same result: energy transfer from $\Eue$
to $\Eli$ necessarily implies a corresponding exhaust of cocurrent
momentum, with the same basic properties, magnitude, and scaling as
derived here. \citep{StoltzfusDueck16rotpop}  

A few comments: $\tmfluxt$ is a residual stress, following from
symmetry-breaking due to energy transfer from $\Eue$ to $\Eli$,
regardless of the background rotation profile. The symmetry breaking
is statistical: it occurs simply because free energy flows through
phase space from sources to sinks. In particular, $\tmfluxt$ is nonlinear,
not quasilinear---it follows from the energy transfer term summed
over all modes (including damped ones) and does not depend on the
linear mode structure of any particular instability. However, it does
require the presence of fluctuations (unstable or damped) at low enough
frequency to excite ion parallel flows, $\om\qq\Mr/\vti\lesssim1$.
It survives in a radially local (fluxtube) limit, not requiring any
radially global effects. Although $\tmfluxt$ results from a higher-order
part of the $\exb$ drift ($\uei^{(2)}$), which should have little
direct impact on the leading-order turbulence, it is slaved to free
energy fluxes that are determined by the leading-order physics. It
can therefore be estimated even by simulations that neglect $\uei^{(2)}$,
simply by evaluating the relevant energy flux term \emph{a posteriori}.

Although quantitative evaluation requires nonlinear simulation, we
may qualitatively compare $\tmfluxt$ with experimental rotation observations.
The general scaling for the countercurrent velocity peaking (roughly
the toroidal velocity at the $\qq=1$ surface minus that at the pedestal
top) is $\fl(\zs\teo/\tio)(\rip/\Lp)\vti\approx5\fl[\teo(\mathrm{keV})/\ip(\mathrm{MA})](\mr/\Lp)\mathrm{km/s}$,
which resembles Rice scaling ($\propto1/\ip$) and has a magnitude
comparable with experimental observations.\citep{deGrassie07,Eriksson09,Rice11nf,Angioni11,McDermott11mom,StoltzfusDueck15rotpop}
Also, in ASDEX-Upgrade (AUG), countercurrent momentum peaking has
correlated strongly with density peaking across many discharge types.\citep{Angioni11}
The relation may be more coincidental than causal: density peaking
tends to occur due to electron precessional resonance for fluctuations
with $\om\lesssim\vti/\Mr$,\citep{Angioni12} which (at core $\qq\sim1$)
are the same modes that can excite ion parallel flows, thus driving
countercurrent peaking. Interestingly, on Alcator C-mod, the presence
of countercurrent peaking is correlated with the disappearance of
broadband high-$\kp$ density fluctuations.\citep{Rice11nf} Viewed
theoretically, dominant dissipation via low-$\kp$ ion parallel flows,
which implies countercurrent rotation peaking in the present model,
would also imply the reduction or elimination of a strong direct cascade
of density fluctuations to high $\kp$, consistent with C-mod measurements.
Further qualitative and quantitative comparisons are needed.

In conclusion, a geometrical correction to the $\exb$ drift causes
an outward flux of cocurrent momentum whenever electrostatic potential
energy is transferred to ion parallel flows. The robust symmetry breaking
follows from the free energy flow in phase space and does not depend
on assumed linear eigenmode structure. The resulting rotation peaking
is countercurrent and scales with $(\zs\teo/\tio)(\rip/\Lp)\vti\propto(\teo/\ip)(\mr/\Lp)$.
This peaking mechanism can only act when fluctuations are low-frequency
enough to excite ion parallel flows, which may explain some recent
experimental observations.\citep{Angioni11,Rice11nf}

\medskip{}

Helpful discussions with C.~Angioni, G.~Hammett, P.~Helander,
J.~Krommes, and B.~Scott, and funding by the Max-Planck/Princeton
Center for Plasma Physics are gratefully acknowledged.

\medskip{}

\bibliographystyle{aipnum4-1}

\end{document}